# Landau Theory of Domain Wall Magnetoelectricity

Maren Daraktchiev[1]*, Gustau Catalan[1,2], and James F. Scott[1,3]

1. *Department of Earth Sciences, University of Cambridge, Downing Street, Cambridge, CB2 3EQ, UK*
2. *ICREA and CIN2 (CSIC-ICN), Campus Universitat Autonoma de Barcelona, Bellaterra 08193, Spain*
3. *Cavendish Laboratory, Department of Physics, University of Cambridge, UK*

We calculate the exact analytical solution to the domain wall properties in a multiferroic system with two order parameters that are coupled bi-quadratically. This is then adapted to the case of a magnetoelectric multiferroic material such as $BiFeO_3$, with a view to examine critically whether the domain walls can account for the enhancement of magnetization reported for thin films fo this material, in view of the correlation between increasing magnetization and increasing volume fraction of domain walls as films become thinner. The present analysis can be generalized to describe a class of magnetoelectric devices based upon domain walls rather than bulk properties.

## 1. Introduction

$BiFeO_3$ was first researched in Leningrad in the 1950s and 60s and has recently had a renaissance due to its room-temperature magnetic and ferroelectric properties [1] - [5]. The report by *Wang* [1] of strong ferromagnetism in thin films of bismuth ferrite ($BiFeO_3$) has stimulated considerable interest in the possibility of using this material as a room-temperature magnetoelectric sensor or memory element. However, later studies have shown that the ferromagnetic moment is very small in other specimens, suggesting that the effect is not an intrinsic bulk symmetry property and may arise from oxygen vacancies, $Fe^{+2}$ ions, or other point or extended defects[6][7][8]. Domain walls are of course a special kind of extended defect, and we show here that they can in principle be ferromagnetic. This is a kind of magnetoelectricity that transcends the bulk symmetry properties and may be large enough to provide a different class of multiferroic device.

Beginning two decades ago, *Lajzerowicz* [9] predicted that domain walls in systems with two coupled order parameters could lead to the emergence of one parameter inside the domain wall of the other. Transposing their results to the specific case of magnetoelectric multiferroics, the centers of domain walls could in principle bear out either a net electric ($P$) or magnetic ($M$) moment while the domain themselves were non-electric ($P_B$=0) or non-magnetic ($M_B$=0). This kind of ferroic behavior was generalized by *Privratska* and *Janovec* [10] in 1997 to show, using group-theoretical arguments, what crystal symmetries might exhibit it. The space group of $BiFeO_3$, R3c, was one of them. Unfortunately, the *Privratska-Janovec* theory could not estimate magnitudes, or even whether the effect could be observable. Later Fiebig and co-workers also studied magnetoelectric effects in the domain walls of antiferromagnetic hexagonal manganites $YMnO_3$ [11] and $HoMnO_3$ [12]. More recently, a linear correlation between the number density of (ferroelastic) domain walls and ferromagnetism has also been reported in thin films of $TbMnO_3$ [13], suggesting that the ferroelastic walls of this antiferromagnetic material may be ferromagnetic. Finally, the converse effect of polarization emerging inside the domain walls of a magnet was beautifully demonstrated by Logginov *et al.* [14].

Taken together, then, all these findings indicate the possibility of having a class of magnetoelectric devices that we shall term "domain wall multiferroics". Although such domain wall effects may be small in bulk, a key point to bear in mind is that thin films have extremely small domains and thus a high volume density of domain walls [15][16][17], so that the domain wall magnetoelectricity may play a very important role in thin film devices. In the present work we provide the analytical solution to the thermodynamic equations that describe the magnetoelectric coupling in the domain walls of a multiferroic system, and use the analysis to examine whether this effect may be behind the reported magnetization of $BiFeO_3$ films.

Applications of free energy models to multiferroic materials necessarily involve three coupled parameters: $M$, the net (weak) ferromagnetic magnetization, which usually arises from canting; $L$, the sublattice magnetization ($L>M\geq0$); and $P$, the ferroelectric polarization. Couplings of all three parameters generally need to be considered, as was done by *Fox et al* [18][19] for multiferroic $BaMnF_4$. The magnetism in that material is a bulk effect directly correlated to the linear coupling of $L$, $M$ and $P$ [18][19][20]. By contrast, in the present paper we try to see whether physical conclusions can be obtained via a much simpler free energy where $L$ is neglected and only biquadratic coupling $P^2M^2$ is considered, the idea being to exact analytical solutions to an approximate model rather than approximate or numerical solutions to an exact model. We emphasize that the biquadratic coupling term is always allowed by symmetry, so it must necessarily exist in all multiferroics. Moreover, since this term is not linear, its magnitude is not limited by the Brown-Hornreich-Shtrikman upper bound [21]. Also, this type of coupling contains implicitly the strain coupling, as electrostrictive and magnetostrictive terms all couple to the square of the order parameter, and therefore it must be important in ferroelastic multiferroics such as the perovskite ones. The analytic solutions enable qualitative and even semi-quantitative predictions that are be compared to the archetypal multiferroic, $BiFeO_3$.

It will be assumed to begin with that the phase transitions can be described by a second order quartic Landau expansion. Though the second order transition is often valid for ferroelectric thin films [22], and has been used to describe thin films of $BiFeO_3$ [23] [24], such approximation is potentially inaccurate for bulk $BiFeO_3$, because its ferroelectric phase transition is first-order [5],[25] [26], which would necessarily require expansion up to at least $P^6$ terms. The analytical solution for the magnetoelectric domain walls in a first order thermodynamic potential ($P^6$ and $M^6$ terms) will therefore be presented also. Although the ferroelectric Landau coefficients have been calculated for thin films [23][24], and estimated also in the appendix of the present paper, the exact measurement of the intrinsic bulk coefficients is nevertheless still pending, and the authors very much encourage the exact measurement of all the coefficients in good quality single crystals. The domain-wall magnetization estimated in this paper does come within the same order of magnitude as the remnant magnetization of $BiFeO_3$ films [1], suggesting that a more precise treatment, using exact values of the Landau coefficients and including the symmetry and sublattice magnetization constraints, could help clarify the long standing issue of the origin of ferromagnetism sometimes reported for this material.

## 2. Second order approximation ($P^4$, $M^4$)

To derive the analytic solutions, we start with the Landau-Ginzburg potential for a magnetoelectric multiferroic with positive bi-quadratic coupling between $P$ and $M$, where both $P$ and $M$ are expanded to quartic terms, the minimum required to describe second order phase transitions.

$$G_{MP} = G_0 + \frac{k}{2}(\nabla P)^2 + \frac{l}{2}(\nabla M)^2 + \Im_{MP}(P,M),$$

with $\Im_{PM}$ defined as (1)

$$\Im_{MP}(P,M) = \frac{a}{2}P^2 + \frac{b}{4}P^4 + \frac{\alpha}{2}M^2 + \frac{\beta}{4}M^4 + \frac{g}{2}P^2M^2$$

where $b > 0$, $\beta > 0$ and $b\beta - g^2 > 0$ to ensure that the non-equilibrium potential is positive definite at high value of $P$ and $M$. The coefficient multiplying the term $P^2M^2$ is an effective parameter that includes both direct coupling between $P$ and $M$ (or $E$ and $H$) and also indirect coupling through strain $s$, consisting of electrostriction $sP^2$ plus magnetostriction $sM^2$. Readers should note that the bi-quadratic coupling $g$ is allowed by all symmetries and, consequently, it will always be present, with bi-linear, linear-quadratic or linear-cubic couplings being also possible for specific point groups [27].

The Curie temperatures of the ferroelectric and ferromagnetic phases do not necessarily have to coincide. In several multiferroics, such as BFO or hexagonal $YMnO_3$, the ferroelectric phase transition ($T_C$) has a bigger critical temperature than the ferromagnetic or antiferromagnetic ($T_{Neel}$) one. For $BiFeO_3$, in particular, the ($P_B \neq 0$, $M_B = 0$) solutions would occur below the Curie temperature ($T_C \sim 820°C$). Since $BiFeO_3$ is an antiferromagnet below the Néel temperature ($T_N \sim 370°C$) we shall extend the ($P_B \neq P_0, M_B = 0$) solution also below the antiferromagnetic ordering temperature ($T < T_N$). The case where the net magnetization is non-zero is also analized later.

The stripe domain walls in $G_{MP}(x)$ are introduced in the z-y plane, so that $M$ and $P$ are one-dimensional functions of variable x only (e.g., $M=M(x)$ and $P=P(x)$), and so is the Landau-Ginzburg potential $G_{MP}=G_{MP}(x)$. Both the domain-wall energy and profile can be obtained by minimizing the energy difference between the system with and without domain wall in an infinite crystal. We shall construct the solutions of (1) in one dimension from the variational minimization of $G_{MP}(x)$ by employing the trial orbits $T(P,M)$ constrained to zero in the $P$-$M$ plane (the order-parameter space). This approach will allow us to obtain classes of analytical solutions of (1) that depend on $T(P,M)$ and $L_{MP}(P,M)$. In fact, our approach is to some extent analogous to the *Rajaraman* [28] and *Montonen* [29] approaches for deriving the analytical solutions of two coupled $f^4$ potentials in field theories. Indeed, starting from

$$0 = \frac{dT(P(x),M(x))}{dx} = \frac{\partial T}{\partial P}\frac{dP}{dx} + \frac{\partial T}{\partial M}\frac{dM}{dx}, \quad (2)$$

together with the soliton solutions of $G_{MP}(x)$,

$$\frac{d^2P(x)}{dx^2} = \frac{1}{k}\frac{\partial \Im_{MP}(P,M)}{\partial P(x)}, \quad \frac{d^2M(x)}{dx^2} = \frac{1}{l}\frac{\partial \Im_{MP}(P,M)}{\partial M(x)}, \quad (3)$$

we have obtained the Rajaraman condition:

$$\left(\frac{\partial T}{\partial P}\right)^2 \int \frac{1}{k}\frac{\partial \Im_{MP}}{\partial P}dP = \left(\frac{\partial T}{\partial M}\right)^2 \int \frac{1}{l}\frac{\partial \Im_{MP}}{\partial M}dM, \quad (4)$$

(4) is an integro-differential equation in terms of $P$ and $M$. The integral has to be evaluated along $T(P,M)$ and the integration constants that are not written explicitly in (4), will vanish. The trial orbits are limited to ones having finite total action only. Periodic orbits or orbits that flow to infinity have not been considered since they have infinite total action [28]. Examples of orbits with finite action, i.e., the energy has to be localized and finite as $x \to \pm \infty$, are the zero-energy orbits that begin at one of local minima and end at another local minima of $\Im_{PM}$ (e.g., another zero of $\Im_{PM}$). The zeros of $\Im_{PM}=0$, i.e., the ones that determine $P(x)$ and $M(x)$ in the middle of domains far away from the center of domain walls, are

$$\begin{aligned} &(0,0), \\ &I. \, (\pm P_0, 0), \\ &II. \, (0, \pm\sqrt{-\frac{\alpha}{\beta}}) \\ &III. \, \left(\pm\sqrt{\frac{\alpha g - a\beta}{b\beta - g^2}}, \pm\sqrt{\frac{ag - \alpha b}{b\beta - g^2}}\right) \end{aligned} \quad (5)$$

where $P_0 = \sqrt{-\frac{a}{b}}$. Note that the minima of $\Im_{PM}$, which are its zeros (see. Eq.5), minimize also the Landau-Ginzburg potential [28].

Readers should note that the model (1) describes second order transitions (i.e., ones associated with continuous changes of order parameters $P$ and $M$ at phase transition temperature) that the system may undergo with temperature from the high temperature (high-symmetry) towards the low-temperature (low-symmetry) phases.

Let us look at $(\pm P_0, 0)$, which corresponds to the low-symmetry phase *I* characterized by a ferroelectric moment only (i.e., a ferroelectric domain wall). The stability condition $\partial^2 \acute{A}_{PM}/\partial P^2 > 0$ in this phase requires that $\gamma > -a/P_0^2$ to allow $(-P_0, 0)$ and $(+P_0, 0)$ to be local minima of $\acute{A}_{MP}$. This condition also provides that the other zeros of $\acute{A}_{MP} = 0$ (i.e. *II* and *III*) do no correspond to any local minima of $\acute{A}_{MP}$. As noted above, *P* and *M* are constrained to go from one local minimum $(-P_0, 0)$ to another minimum $(+P_0, 0)$. A zero-energy orbit for $(\pm P_0, 0)$ would be, for example, $0 \equiv T(P,M) = f(P^2 - P_0^2) + M^2$, which together with (3) and (4), determines the *P* (kink) – *M* (breather) domain wall solution that minimizes both $\acute{A}_{MP}$ and $G_{MP}$. The kink-breather solution a topological soliton (topological number different from zero) derived firstly in the field theory [28][29]. Only a limited number of zero-energy orbits are found to minimize (1) and satisfy (4) simultaneously. Solutions constrained on orbits as $T(P,M)=f(P^2-P_0^2)+M^r$, where $r>2$, are excluded, as they do not satisfy (4) or minimize (1). Readers should note that *all* possible domain-wall profiles minimizing (1) depend on the degree of complexity of $T(P,M)$, and therefore cannot be derived explicitly from (4). Kink-breather domain wall profiles satisfying (4) have been systematically calculated from (1) for all tested zero-energy orbits that start at $(-P_0, 0)$ and end at $(+P_0, 0)$.

In the ferroelectric phase *I*, very different scenarios take place depending on the relative signs and magnitudes of the coefficient *a* that multiplies $M^2$, and the magnetoelectric coupling coefficient *g*. It is easy to visualize this by considering only the magnetic and magnetoelectric parts of the free energy, $\Im_M(M) = \frac{a}{2}M^2 + \frac{b}{4}M^4 + \frac{g}{2}P^2 M^2$. The terms multiplying $M^2$ may be regrouped as:

$$\Im_M(M) = \left(\frac{a}{2} + \frac{g}{2}P^2\right)M^2 + \frac{b}{4}M^4 \qquad (6)$$

From this re-grouping, it emerges that the effect of the magnetoelectric coupling is a renormalization of the lowest-order magnetic coefficient, and hence of the magnetic transition. Assuming a positive sign for $\gamma$, and given that $P\neq0$ in the ferroelectric state, the magnetoelectric coupling makes the magnetic coefficient "more positive". If $a>0$ (i.e., if $T>T_{Neel}$), this is of no consequence whatsoever. But, if $a<0$ ($T<T_{Neel}$), then something very interesting may happen. Specifically, if $a<0$ (i.e., if the material is below $T_{Neel}$) but the absolute value of the coefficients is such that $\frac{a}{2} + \frac{g}{2}P^2 \geq 0$, then the material will still display no magnetism. *Except, of course, where the polarization is suppressed, i.e., inside the domain wall.* In the middle of the domain wall, the polarization is zero and hence the magnetoelectric coupling is suppressed, so that the renormalization is cancelled and magnetism can emerge. The analytic expressions for $P(x)$ and $M(x)$ in this scenario, calculated from the Landau-Ginzburg potential (1), are

$$P(x) = P_0 \tanh\left(\frac{x}{d_{MP}}\right), M(x) = \pm M_0 \operatorname{sech}\left(\frac{x}{d_{MP}}\right), \qquad (7)$$

where $M_0$ is the magnetization in the center of the domain wall (see Fig. 1b) and $P_0$ is the polarization in the center of the domains (see Fig. 1a)

$$P_0 \equiv \sqrt{-\frac{a}{b}}, \quad M_0 \equiv \sqrt{\frac{ga - 2ab}{bb}}, \qquad (8)$$

Where a<0 (ferroelectric state). $P(x)$ and $M(x)$ are constrained on an ellipse $M^2/M_0^2 + P^2/P_0^2 = 1$ in the *P-M* plane (see Fig. 1c). A similar relationship between the order parameters has been obtained in the spherical model and its relaxor relative, the spherical random-bond, random field model [32]. The kink-breather domain-wall profile has the homotopic invariant, $I(P,M) \equiv \int dx(P'\partial M/\partial x)$ equals in magnitude to the orbit area. As $I(P,+M) = -I(P,-M)$ therefore, the kink-breather domain-wall profile would possess total magnetoelectric chirality different from zero depending on which branch of the ellipse *M* travels along (Fig. 1c). This result complements *Lajzerowicz's* [9] who predicted a similar effect in ferroelastic domain walls. These authors also showed that the existence of coupling between two ferroic order parameters may lead to chirality, which account for the observation that the ferroelectric domain walls of BFO are Heisenberg-like rather than Ising-like, that is, the ferroelectric polarization rotates across the wall [33]. Incidentally, the existence of chirality inside 180° walls has also been proposed for non-magnetoelectric ferroelectrics such as $BaTiO_3$ [34]: although these are not magnetic, however, they are still multiferroic as they are ferroelectric and ferroelastic, so that the chirality predictions of Lajzerowicz [9] may indeed be applied to it.

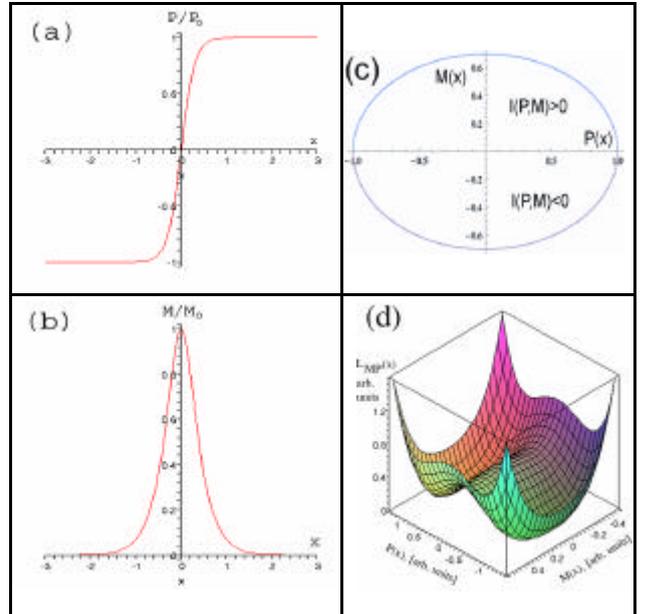

FIG 1 Polarization-magnetization domain-wall profile when domains are not ferromagnetic: (a) P≡0 in the center of domain walls but P≠0 in the domains; (b) M≠0 in the center of domain walls but M≡0 in the domains; (c) $M(x)$ and $P(x)$ are constrained on an ellipse; (d) $\acute{A}_{MP}(x)$ shows two minima at $P=\pm P_0$. The system will move from one of those minima to other through a maximum value at 0. This will gives an increase of *M* in the center of domain walls. x-Axes in (a), (b) and (d) are in arbitrary units.

Eq. (7-8) shows that a net magnetization can appear in the domain walls due to the magnetoelectric coupling and the gradient terms, even when the domains themselves have no net magnetization. Similarly, *Goltsev and Lottermoser* [11][12] have reported that antiferromagnetic multiferroics, such as *YMnO_3* and *HoMnO_3*, can also reveal intrinsic net magnetization in the center of the domain walls due to magnetoelectric coupling. Our result agrees with concepts discussed in [9]-[12] and extends them to the analytical solutions of spatial variation of magnetization and polarization to the entire width of the domain walls as functions of domain-wall thickness. The domain wall thickness is given by

$$d_{MP} \equiv 2^{\frac{1}{2}} \sqrt{\left(\frac{bbk}{2gba - ag^2 - abb}\right)} \quad (9)$$

and the Landau coefficient $k>0$, $l>0$, $\alpha<0$, $\beta>0$, $a<0$, $b>0$, $g>0$. Eq. (9) shows that the ferroelectric domain walls $d_{MP}$ in magnetoelectrics will *always* be wider than the ferroelectric ones $d_P=2^{1/2}(-k/a)^{1/2}$ in a pure ferroelectric. These results are only fulfilled at condition $\gamma > ab/a$, which states explicitly that at the low-temperature phase $(\pm P_0, 0)$ the coupling has to be *strong enough* in the quartic Landau expansion to allow the magnetization to come into existence at the middle of domain walls. Yet another scenario takes place when $a<0$ and $\frac{a}{2}+\frac{g}{2}P^2 \leq 0$. In this case, the material has net magnetism both inside and outside the domains but, crucially, as shown by eq. (6), the coefficient multiplying the $M^2$ term is more negative (and hence the magnetism bigger) inside the domain walls than outside, due to the suppression of the magnetoelectric coupling. Hence, an increase over the bulk magnetism may be expected at the walls also when the material is aleady in a magnetic state. This situation may apply to bismuth ferrite thin films, which are known to have a canting moment below $T_{Neel}$. Because BFO has a first order phase transition, though, this situation is best analyzed using an LGD expansion up to 6th order terms.

### 3. First-order ferroelectric transition, second-order magnetoelectric transition

Strictly speaking, the model (1) should not be applicable to bismuth ferrite, as the ferroelectric transition in this material is found experimentally to be first-order [5], which requires the ferroelectric order parameter to be expanded up to order $P^6$ –although the approximation to $P^4$ terms has nevertheless been used to describe strain effects in thin films of this material [23]. *Toledano* [31], *Gufan* [35] and *Holakovsky* [36] have shown that the model depicted by Eq. 1 is unsuitable for negative interactions between the order parameters ($g<-\sqrt{bb}$), which may limit the model's applicability in the description of experimental results in the $(\pm P_0, 0)$ phase. By including $P^6$ and $M^6$ terms in (1) we may show that the kink-breather solution (eq. 7) is again a solution of the amended potential and overcomes the constraint on $\gamma$ to be large in the $(P_B \neq 0, M_B=0)$ phase in order to induce domain wall magnetization in the quartic Landau model. Thus, we consider a potential expended up to $P^6$ and $M^6$ terms:

$$G_{MP} = G_0 + \frac{k}{2}(\nabla P)^2 + \frac{l}{2}(\nabla M)^2 + \Im_{MP}(P,M) \quad (10)$$

with a potential $\Im_{PM}$ defined as

$$\Im_{MP}(P,M) = G_0 + \frac{k}{2}(\nabla P)^2 + \frac{l}{2}(\nabla M)^2 + \frac{a}{2}P^2 + \frac{b}{4}P^4 + \frac{h}{6}P^6 + \frac{a}{2}M^2 + \frac{b}{4}M^4 + \frac{n}{6}M^6 + \frac{g}{2}P^2M^2.$$

$n > 0$ and $h > 0$ insure that $G_{MP}$ is positive definite at high value of $P$ and $M$. Some zeros of $\Im_{PM}$ are

$$(\pm P_0, 0),$$
$$\left(0, \pm\sqrt{\frac{-b+\sqrt{b^2-4an}}{2n}}\right) \quad (11)$$

where $P_0 = \sqrt{\frac{-b+\sqrt{b^2-4ah}}{2h}}$.

In the low-symmetry ($P_B\neq 0$, $M_B\neq 0$,) phase, equivalent to *III* in (5), $M_B$ and $P_B$ are solutions of the following polynomial equations to the eighth power:
$M_B$ = RootOf[$\eta n^2 x^8 + 2b\eta n x^6 + (-\beta\gamma n + \eta b^2 - 2\eta na)x^4 + (-\beta\gamma b - 2\eta ab + \gamma^3)x^2 - \alpha\gamma^2 + \beta\gamma b + \eta a^2$]
$P_B = (a - bM_B^2 - dM_B^4)^{1/2}/\gamma^{1/2}$.

### 3.1 Ferroelectric, non-magnetic domains, $P\neq 0$, M=0

In (11), we look at the case $a<0$ and $a>0$ in the intermediate phase $T_{Neel}<T<T_C$, which states explicitly that the domains in *BiFeO_3* are in the ferroelectric state of polarization $P_0$ above the magnetic ordering $(M_B=0)$ temperature. Though $g$ is normally positive, it can be small or even slightly negative in (10). We again consider a zero-energy orbit $0 \equiv T(P,M) = f(P^2-P_0^2) + M^2$ for $(-P_0,0)$ and $(+P_0,0)$ minima to calculate the domain wall profiles that minimize (10) and satisfy (4) simultaneously. The analytic expressions for $P(x)$ and $M(x)$ across the domain walls in the stability field of low-symmetry $(P_B=\pm P_0, M_B=0)$ phase are

$$M(x) \equiv \frac{\sqrt{2c_1}}{\sqrt{\sqrt{c_2^2 - 3c_1c_3}\cosh(2\sqrt{c_1}x)+c_2}}$$

$$c_1 = \frac{a+gP_0^2}{l} \quad (12)$$

$$c_2 = \frac{(16nP_0^8 h^2 + 24nP_0^6 hb + 9nP_0^4 b^2 + 27g^3 P_0^2 + 27g^2 a)}{6gP_0^2 l(4hP_0^2 + 3b)}$$

$$c_3 = \frac{n}{l}$$

and

$$P(x) \equiv \sqrt{P_0^2 - \frac{3g}{4hP_0^2 + 3b}M(x)^2}, \quad (13)$$

where $\alpha<0$, $a>0$, $\gamma\geq 0$, and

$$d_{MP} \equiv \frac{1}{2\sqrt{c_1}} = \sqrt{\frac{l}{4(a + gP_0^2)}}. \qquad (14)$$

Studying (12)-(13) one observes that they have standard forms of bell-shaped $M(x)$ and kink-shaped $P(x)$ as those in Fig. 1. $M(x)$ and $P(x)$ are again constrained on an ellipse. The limit $g \to 0$ forces $P(x) \to P_0$ and $M(x) \to 0$ in (11)-(12), a result that is intuitively expected to be correct when $M(x)$ and $P(x)$ are decoupled in the Landau-Ginzburg potential. A large $g >> 0$ would induce a dramatic increase of $M(x)$ in the middle of domain walls when $P(x)$ inverts through zero (see Fig. 1), while, $g \to 0$ will not induce any magnetic moments within domain walls with inversion of $P(x)$ at zero, and thus the material will remain non magnetic both within the walls and, of course, within the domains.

Building on (12), (13) and (14), and using the Landau coefficients estimated in the Appendix, we calculate $d_{MP} = 6 \times 10^{-9}$ m, $M_0 = 1.3 \times 10^5$ A/m = 130 emu/cm$^3$ for $BiFeO_3$ (see Fig. 2b). This magnetization inside the domain wall is very large, as it represents nearly 1 Bohr magnetron per unit cell. It is also noteworthy that the expected ferroelectric domain wall thickness of BFO is considerably thicker than that of "pure" ferroelectrics, and closer to that of magnetic domain walls. Specifically, the domain wall thickness D=2$d_{MP}$=12nm is comparable to that of pure ferromagnets with high anisotropy, such as cobalt. This result had been anticipated on the basis of the big domain periodicity observed in BFO thin films [15] and the known proportionality between domain size and domain wall thickness [16].

The remnant magnetization of *Wang's* samples [1] for both the in-plane and the out-of-plane loops measured on 70-nm-thick $BiFeO_3$ film by vibrating sample magnetometry is $M_{remnant} \sim 10\text{-}15$ emu/cm$^3$, with the reported saturation magnetization being considerably bigger, around 150 emu/cm$^3$ (i.e., almost 1$\mu_b$ per unit cell), although more recent measurements give a lower value around 60 emu/cm$^3$ [44]. Our intention is to quantify whether this experimental $M_{remnant}$ may be caused by the domain wall magnetization. For this, we approximate the average contribution of domain wall magnetization to the macroscopic magnetization of the film as $M_{DW} \approx M_0 d_{MP}/w = M_0 \times 2 d_{MP}/w$, where the ratio $d_{MP}/w$ measures the volume fraction of domain walls in the film. Typical values for the stripe domain widths in the thin films are in the region $w \sim 100 \times 10^{-9}\text{-}400 \times 10^{-9}$ m [45], with fractal domains being considerably smaller [15], so that their domain wall contribution may be larger. Using $d_{MP} = 6 \times 10^{-9}$ m, w=200nm and $M_0 = 1.3 \times 10^5$ A/m in the centre of the walls this leads to a net macroscopic magnetization of $M_{DW} \sim 7800$ A/m $\sim 8$ emu/cm$^3$, which is reasonably close to the experimental remnant magnetization given the crudeness of some of our approximations. This, of course neglects the already existing canting moment of $BiFeO_3$, which is also of the order of 6-8 emu/cm$^3$. If we simply add the two, we obtain a net magnetization of the order of 14-16 emu/cm$^3$, which is close to the value experimentally measured.

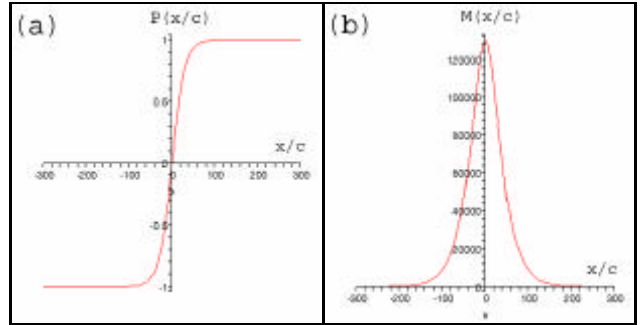

FIG 2 (a-b) Polarization-magnetization domain-wall profile in BeFeO$_3$; $M(x/c)$ and $P(x/c)$ are constrained on an ellipse (12). X Axis is normalized to x/c, where c = $4 \times 10^{-10}$ m is the lattice constant. The system will move from $P_0$=-1 to $P_0$=+1, giving an increase of $M$ in the center of domain walls. The Landau coefficients are given in the appendix.

Our results therefore show that domain walls could in principle account for a significant fraction of the enhancement in remnant magnetization in BFO thin films, particularly for films with fractal walls, which have smaller domains and a larger perimeter-to-area ratios (i.e., bigger ratios of wall to domain). The domain wall magnetization, on the other hand, is still too small to explain the large measured saturation magnetization, although we emphasize that our calculations are done only for zero external field and do not really attempt to address what happens at saturation fields. Of course, our solution does not exclude further contributions from any of the other explanations already put forward in the literature such as vacancy-induced $Fe^{2+}$ oxidation state [6], parasitic ferromagnetic phases [7], or localized "hotspots" of magnetite generated upon electric field cycling [8].

It is also worth mentioning that domain walls are known to attract charge carriers and oxygen vacancies [41], [42], [43]. These could further enhance the local magnetization at the domain wall beyond the value calculated here. Thus, domain walls may contribute to the magnetization both intrinsically, as calculated above, or extrinsically via concentration of oxygen vacancies in their interior. This latter mechanism would also provide an explanation for the fact that, while BFO magnetization is known to be linked to the O pressure during fabrication, the level of O vacancies required to explain the magnetization is too high to be compatible with the pristine structure observed by x-ray diffraction; if O vacancies were not in the domains but localized inside the narrow domain walls, they would avoid being observed by macroscopically averaged techniques such as x-ray diffraction while still contributing to strong localized magnetization.

On the other hand, other groups have not observed any enhancement in remnant magnetization [7] in spite of the fact that their films clearly possess domain walls [15]. Two explanations may be invoked here. Firstly, the type of domain wall is important. Our analysis is for 180º domain walls, i.e., those where the polarization is completely reversed. BFO, however, is known to have also ferroelectric/ferroelastic domain walls where the polarization rotates by approximately ~109º (separating domains where only two of the cartesian components of the

111-oriented polarization are reversed) and ~71° (those where only one component is reversed). Naively one may expect the magnetization in these walls to be a fraction of that in the 180° wall; if the magnetization is only enhanced for the fraction of polarization that is reversed, then the 109° walls will have smaller magnetization than the 180 walls, but bigger than 71° walls. In this respect, it is noteworthy that Ramesh *et al.* have noticed that films with only 71° walls have no appreciable enhancement of magnetization, in contrast with those with 109° walls which do [44]. Since different substrate orientations, deposition rates and electrode configurations favor different type of domains, it is conceivable that the samples from different laboratories might have different predominance of one type or another of domain wall, thereby accounting for the different magnetic behaviour. A detailed study of the correlation between domain type and magnetization is needed in order to clarify this point.

### 3.2. Ferroelectric and magnetic domains

While bulk BFO is an antiferromagnet with zero net magnetization on account of the cycloidal spin rotation, in thin films the spin cycloid is destroyed. This leads to weakly canted spin configuration with weak ferromagnetism (ferrimagnetism) which can be further enhanced by impurities or doping. Accordingly, it is perhaps inappropriate to treat BFO thin films as having strictly zero net magnetization in zero field, as done above, and instead one should treat it as a material which is both ferroelectric and (weakly) ferromagnetic; in mathematical terms of the Landau theory, one where both α and $a$ (the first-order coefficients of the polarization and the magnetization) are negative. Therefore, $M$ would be non-zero in the domains, $M_B \ne 0$.

The solutions for ferroelectric-ferromagnetic domain walls are derived analytically from (1-2) by constraining them on a zero-energy orbits that begin and end at

$$P_B = \pm\sqrt{\frac{a\mathbf{g}-a\mathbf{b}}{\mathbf{b}\mathbf{b}-\mathbf{g}^2}}, \quad M_B = \pm\sqrt{\frac{a\mathbf{g}-a\mathbf{b}}{\mathbf{b}\mathbf{b}-\mathbf{g}^2}}, \qquad (15)$$

$a < 0$ and $a < 0$ in (14). Those are solutions of the low-symmetry phase characterized by both ferroelectric and ferromagnetic moments. The stability conditions $\partial^2 Á_{PM}/\partial P^2 > 0$, $\partial^2 Á_{PM}/\partial M^2 > 0$ and $\partial^2 Á_{PM}/\partial M/\partial P > 0$ in this phase provide that $\gamma < a\mathbf{b}/\mathbf{a}$ or $\gamma < a\mathbf{b}/\mathbf{a}$, meaning that $\gamma$ has to be small enough to allow $(-P_B, -M_B,)$, $(-P_B, +M_B,)$, $(+P_B, -M_B,)$ and $(+P_B, +M_B)$ to be local minima of $Á_{MP}$ (see Fig. 3d). This condition also provides that the other zeros of $Á_{MP}=0$, i.e. $(-P_0, 0)$ and $(+P_0, 0)$, do no correspond to any local minima of $Á_{MP}$. The potential $Á_{MP}$ will have four zeros (see Fig. 3d) that define all possible orientations (up- and down-direction) of $P$ and $M$ moments in domains. This contrasts the potential at Fig.1d which has two zeros with two possible orientation of $P$ (up and down) in domains.

#### 3.2.1. Walls that invert P and M simultaneously

The variational minimization of (1) constrained on the zero-energy orbit $0 \equiv T(P,M) = f(P^2-P_B^2) + Â(M^2-M_B^2)$, together with (3) and (4), determines the $P$ (kink) – $M$ (kink) domain wall solution that minimizes $Á_{MP}$ and $G_{MP}$ in (1). Similarly to the kink-breather solution discussed above (see Fig. 1 and 2), the kink-kink solution is also a topological soliton solution studied firstly in the field theory [28], [29]. Analytical expressions for $P(x)$ and $M(x)$ across the domain walls in the stability field of low-symmetry ($P_B$, $M_B$) phase are either (see Fig. 3)

$$P(x) = P_B \tanh\left(x/d_B\right), \quad M(x) = \pm M_B \tanh\left(x/d_B\right), \qquad (16)$$

where

$$d_B = 2^{\frac{1}{2}}\sqrt{\frac{k}{\mathbf{g}M_B^2 + \mathbf{b}P_B^2}} = 2^{\frac{1}{2}}\sqrt{-\frac{k}{\mathbf{a}}} \equiv d_P, \qquad (17)$$

or

$$P(x) = \sqrt{\frac{a}{l}\frac{(l\mathbf{g}-k\mathbf{b})}{(\mathbf{b}\mathbf{b}-\mathbf{g}^2)}}\tanh\left(x/d_B\right), \qquad (18)$$

$$M(x) = \pm\sqrt{-\frac{a}{l}\frac{(l\mathbf{b}-k\mathbf{g})}{(\mathbf{b}\mathbf{b}-\mathbf{g}^2)}}\tanh\left(x/d_B\right),$$

where

$$d_B = 2^{\frac{1}{2}}\sqrt{-\frac{l}{\mathbf{a}}} \equiv d_M \qquad (19)$$

In (16) - (19), $\mathbf{a}<0$ and $a<0$.

The above results reveal that the domain-wall thicknesses in the ferroelectric-magnetic $(\pm P_B, \pm M_B)$ phase in BFO would be as as thick as those in pure ferromagnets. However, the reader should note that the domain walls (16) - (19) would not contribute to the remnant magnetization as $P(0)=0$, $M(0)=0$.

$P(x)$ and $M(x)$ are constrained on a line with equation

$$M(x) = \mp\frac{\sqrt{(a\mathbf{b}-\mathbf{g}a)(a\mathbf{b}-\mathbf{g}a)}}{\mathbf{g}a-a\mathbf{b}}P(x), \qquad (20)$$

being true for $P(x)$ and $M(x)$ from (15),

$$M(x) = \pm\frac{\sqrt{(k\mathbf{b}-l\mathbf{g})(l\mathbf{b}-k\mathbf{g})}}{k\mathbf{b}-l\mathbf{g}}P(x) \qquad (21)$$

or $P(x)$ and $M(x)$ from (19) in the $P$-$M$ plane (see Fig. 3c, which is an example drawn for (16) only).

#### 3.2.2. Walls that invert polarization but not magnetization

Importantly for our purpose, $P(x)$ and $M(x)$ across the domain walls in the stability field of low-symmetry ($P_B$, $M_B$) phase can also obey the following relations (see Fig. 4)

$$P(x) = P_C \tanh\left(x/d_C\right), \qquad (22)$$

$$M(x) = M_C\sqrt{2-\tanh\left(x/d_C\right)^2},$$

where

$$P_C = \pm \frac{\sqrt{2}}{2} \sqrt{\frac{2bbP_B^2 + gbM_B^2 - g^2 P_B^2}{bb}}, \quad (23)$$

$$M_C = \pm \frac{\sqrt{2}}{2} \sqrt{\frac{gP_B^2 + bM_B^2}{b}},$$

and

$$d_C = \frac{2}{P_B} \sqrt{\frac{kb}{(bb - g^2)}} \quad (24)$$

with $a<0$, $a<0$.

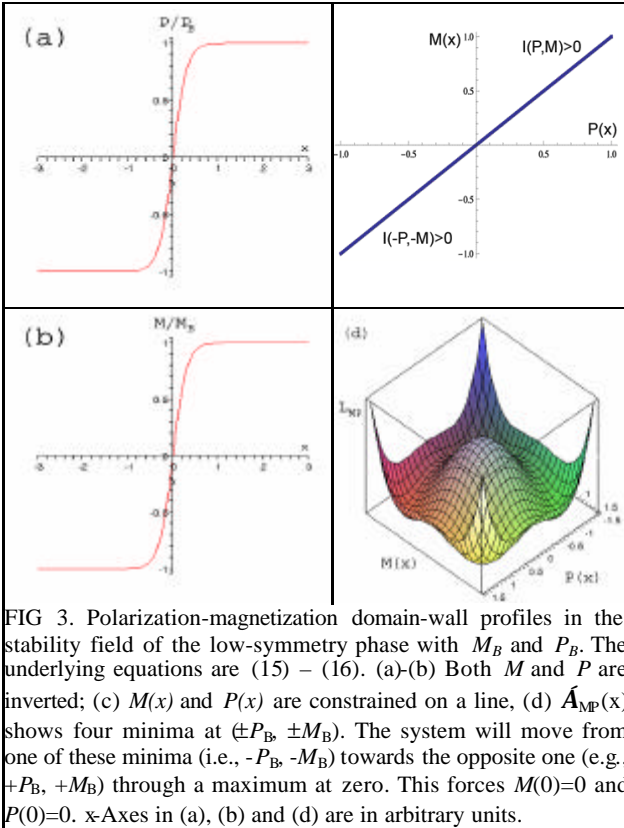

FIG 3. Polarization-magnetization domain-wall profiles in the stability field of the low-symmetry phase with $M_B$ and $P_B$. The underlying equations are (15) – (16). (a)-(b) Both $M$ and $P$ are inverted; (c) $M(x)$ and $P(x)$ are constrained on a line, (d) $Á_{MP}(x)$ shows four minima at $(\pm P_B, \pm M_B)$. The system will move from one of these minima (i.e., $-P_B, -M_B$) towards the opposite one (e.g., $+P_B, +M_B$) through a maximum at zero. This forces $M(0)=0$ and $P(0)=0$. x-Axes in (a), (b) and (d) are in arbitrary units.

This type of domain wall corresponds to an inversion of ferroelectric polarization without inversion of magnetic moment. Such walls would be thicker than one in pure ferroelectrics and, by contrast to the scenario depicted by (16)-(19), here the domain walls would contribute to an enhanced remnant magnetization as $P(0)=0$, $M(0) = \sqrt{-a/b}$.

$P(x)$ and $M(x)$ are constrained on

$$\frac{M(x)^2}{M_C^2} + \frac{P(x)^2}{P_C^2} = 2 \quad (25)$$

showing that domain-wall profile would also possess magnetoelectric chirality. The profile of $Á_{MP}$ is similar to that drawn in Fig. 1d with two local minima and therefore, will not be plotted in Fig. 4. Fig. 1d reveals that the system moves from one of $Á_{MP}$ minima towards another minimum without reverting $M(x)$ (i.e., bell-shaped $M(x)$ in Fig. 4b). However, $P(x)$ will invert at zero giving $P(0)=0$ (i.e., kink-shaped $P(x)$ in Fig. 4a). Furthermore, $M = \sqrt{2}M_B$ in the center of domain walls while $M=M_B$ in the domains (see Fig. 4b). (22) so that the domain wall magnetization is a fraction of the total macroscopic magnetization in the low-symmetry ($P_B$, $M_B$) phase (see Fig. 4b), which is dramatically different from the scenario in the low-symmetry ($P_0$, 0) phase (see Fig. 1b), where the domain walls carry out the total magnetization of $BiFeO_3$ as the stripe domains are not magnetic but ferroelectric (see Fig. 1a).

Eq. 22 shows that the maximum enhancement of magnetization that may be expected in the latter case is M(wall)=$\sqrt{2}M(domain)$. Given that the canting moment of BFO is 8 emu/cm$^3$, that would lead to a maximum moment of ~ 11 emu/cm$^3$ inside the wall, in excellent agreement with recent ab-initio calculations by Lubk *et al*.[48], who predict a 33% increase in the canting angle, suggesting that the intrinsic magnetization of the wall may indeed be larger than that of the domain. Nevertheless, once averaged by the volume fraction of domain wall material, this enhancement is too small to account by itself for the total magnetic moment observed in thin films.

### 4. Conclusions

The present paper was intended to stimulate further work on multiferroics such as BiFeO$_3$ and to suggest a mechanism for weak ferromagnetism in thin films that might reconcile previous disparate results. Although it is not justified to extract exact numerical conclusions in the case of BiFeO$_3$ from our simplified free energy, it is nevertheless encouraging to show that plausible results are obtained which are quantitatively within the right order of magnitude. This may provide motivation for more realistic free energy models in the future, as in [18][19], once more detailed data are obtained for bismuth ferrite. However, we stress that using a more complicated free energy, as done in [18][19], introduces many unknown parameters and is generally intractable analytically, so that only numerical methods can be deployed.

Since thickness, strain, grain size and, above all, impurity levels, can strongly affect to what extent the BFO films are antiferromagnetic or weakly ferromagnetic to start with, they will also affect which of the different discussed scenarios is relevant and thus also whether the domain wall magnetization is the key issue in the first place. Likewise, whether the transition can be described by a second order potential (as in epitaxial thin films) or first order (as in bulk), or even some intermediate, tricritical state (not considered here) also has an effect on the predicted behaviour of the magnetization. Once again, the specific circumstances of the samples are critical and may explain the difference both within and between samples made in different laboratories. As the main take-home messages, then, we would emphasize that (i) domain wall magnetization can be big enough to measurably increase the average magnetic moment of thin films and (ii) whether or not they account for the magnetization of specific samples will depend strongly on the specific characteristics of the samples.

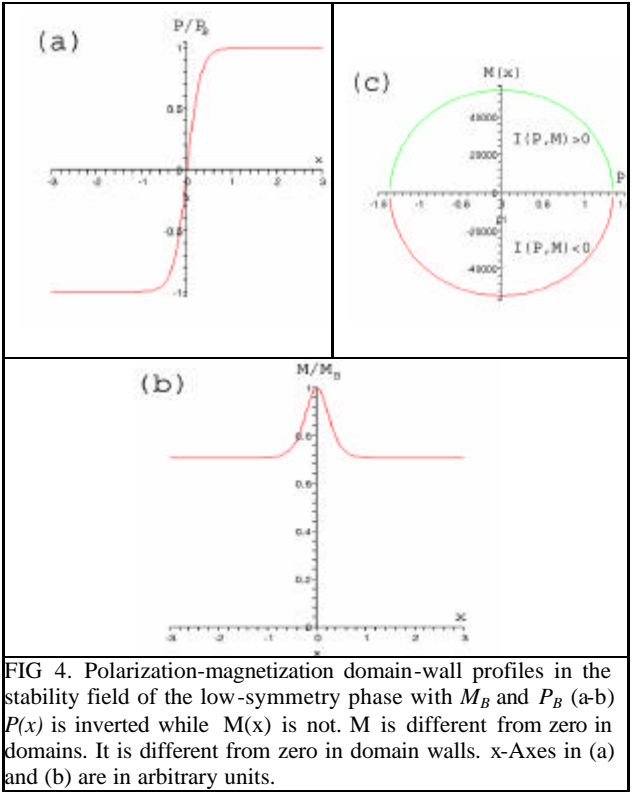

FIG 4. Polarization-magnetization domain-wall profiles in the stability field of the low-symmetry phase with $M_B$ and $P_B$ (a-b) $P(x)$ is inverted while $M(x)$ is not. M is different from zero in domains. It is different from zero in domain walls. x-Axes in (a) and (b) are in arbitrary units.

**Appendix: Landau Coefficients of BiFeO$_3$**

We provide below estimates of the Landau coefficients for BFO. The minimization of $G_{MP}$ with respect to $P$ yields $a+bP^2+hP^4+gM^2=0$. Then, the equilibrium value of $P$ in the bulk yields

$$P^2 = \frac{-b+\sqrt{b^2-4ha-4hgM^2}}{2h}. \quad [A1]$$

The inverse permittivity is the second derivative with respect to the polarization and in the ferroelectric phase it is

$$c_l^{-1}=-a+2bP+4hP^3+gM^2. \quad [A2]$$

In order to calculate the coefficients in these expressions, we must compare with experimental results. In order to facilitate this comparison, we shall make a series of assumptions, and discuss their validity.

Firstly, although the ferroelectric phase transition in bulk BFO is first order [5][25][26], it is common for the transitions of ferroelectrics to become second order when the samples are epitaxially clamped thin films [22]. In these conditions, the effective coefficient of the $P^4$ term, $b$, is positive, while $hP^6$, which is much smaller in absolute value, can be discarded. The second order approximation has been used before to describe thin films of BiFeO$_3$ [23][24]. The second approximation that we shall make is that the biquadratic magnetoelectric coupling is relatively small, so that it can be treated as a perturbative contribution to the bare ferroelectric solution.

Starting with equation [A1], and making the simplifications mentioned above (second order approach, small magnetoelectric coupling), the polarization of a simple ferroelectric is $P_0^2 = -\frac{a}{b}$, while the solution including the magnetoelectric coupling is $P^2 \approx P_0^2 - \frac{g}{b}M^2$. The magnetoelectric polarization is usually much smaller than the ferroelectric one, so a first-order Taylor expansion yields $P \approx P_0 - M^2 g/(2bP_0)$. Meanwhile, the inverse ferroelectric permittivity is (eq. A2) $c_l^{-1}=-a+2bP+gM^2 \approx 2bP_0^2$. Finally, the inverse magnetic susceptibility is found by differentiating the free energy with respect to M. For the phase ($P=P_0$, $M=0$) this is $c_M^{-1}=a+gP_0^2 \approx a$. Thus, the $a$, $b$, $g$ and $a$ coefficients can be extracted from the bulk permittivity, susceptibility and polarization.

The dielectric constant of BiFeO$_3$ at room temperature is $c_f \sim 30\varepsilon_0$ [25] and the polarization of good quality single crystals is around $P_0 \sim 1$ Cm$^{-2}$,[37][38], hence $a \sim -1.7 \times 10^9$ C$^{-2}$m$^2$N, $b \sim 1.7 \times 10^9$ C$^{-4}$m$^6$N (S.I). The magnetic susceptibility of BiFeO$_3$ is $c_M = 0.6 \times 10^{-5}$ g$^{-1}$emuOe$^{-1}$ = $4.7 \times 10^{-5}$ C$^{-2}$mkg, so that the first order Landau coefficient is a$\sim 2.1 \times 10^4$ C$^2$m$^{-1}$kg$^{-1}$.

As for the magnetoelectric coefficient, simple arithmetic shows that $P$ (i.e., $P = P_0 - M^2 g/(2bP_0)$) expressed in terms of $c_l^{-1}$ (i.e., $c_l^{-1} = 2bP_0^2$) is $P=P_0+gc_fP_0M^2$ and the magnetoelectric polarization is therefore, $P_{ME} \approx gc_fP_0M^2$. BiFeO$_3$ has no net magnetization in the absence of magnetic fields (the small canting moment is cancelled by a long-period spin cycloid), and the magnetization is essentially proportional to the external magnetic field, so $M=c_MH$ and $P_{ME} \approx gc_fP_0c_M^2H^2$. Thus, the magnetoelectric polarization can be calculated from the $P$-vs-$H$ curves reported in [39]. From these, one finds $P_{ME} \approx \zeta H^2$ for low magnetic fields, where $\zeta = 4.4 \times 10^{-8}$ Cm$^{-2}$T$^{-2}$ = $7 \times 10^{-20}$ CA$^{-2}$. Supplying these values in $\zeta$ allows us to calculate the magnetoelectric coupling coefficient $\gamma=\zeta\chi_f^{-1}\chi_M^{-2}P_0^{-1}$ = 0.12 mkg$^{-1}$.

We also need to calculate the gradient energy coefficients, $l$ and $k$. Referring to [40] and the value of $\chi_M^{-1}$ reported above, we get $\lambda = 3 \times 10^{-12}$ C$^2$kg$^{-1}$m. The value of the ferroelectric "exchange constant" $k$ is not known exactly for BiFeO$_3$, but we assume a value of $k \sim 1 \times 10^{-10}$ C$^{-2}$m$^5$kgs$^{-2}$, which is typical for other perovskite ferroelectrics, although the very high Tc of BFO probably means that this is probably an underestimate ($k$ is a measure of the dipole-dipole interaction, which is expected to be stronger in ferroelectrics that have high ordering temperatures).

Other sources of uncertainty are the value of $a \sim c_M^{-1}$, which has been calculated by neglecting the higher-order terms in the magnetic susceptibility, and the value of $h$. For the latter, the only value reported in the literature is $h=6 \times 10^6$ m$^9$C$^{-4}$F (S.I.),. Note that this is approximately three orders of magnitude lower than $b^2/4/a$ in our set of coefficients, but the difference is smaller if comparing to the other coefficients in [24]. If $h$ were bigger than $b^2/4a$ this could double or triple the value of $M_0$ (e.g., $M_0$=4829.972 A/m and $M_{DW}$=20 emu/cm$^3$), so that the $M_0$ calculated using these coefficients ($M_0 = 1.3 \times 10^5$ A/m, $M_{DW} \sim 7800$ A/m) can in fact be considered as a conservative estimate. The complete

set of Landau coefficients used for our calculations is summarized in the table below (S. I. units):

| α | β | η | γ | κ | λ | a |
|---|---|---|---|---|---|---|
| $-1.7 \times 10^9$ | $1.7 \times 10^9$ | $6 \times 10^6$ | $0.12$ | $1 \times 10^{-10}$ | $3 \times 10^{-12}$ | $2.1 \times 10^4$ |

------------------------

[*]Electronic addresses:
Daraktchiev (Maren.Daraktchiev@dh.gsi.gov.uk), Catalan (gustau.catalan@cin2.es), Scott (jfs32@hermes.cam.ac.uk).